# The lifting effect of an effervescent tablet

Anastasios Koimas

Teacher of Physics at Secondary Education, 1o Gymnasium of Agios Dimitrios, Greece

E-mail: tkoimas@sch.gr

*Keywords: buoyancy , effervescent tablet , surface to volume ratio , mathematical toy model , educational experiment*

**Abstract**
The lifting effect of an effervescent tablet in the water is ideal for educational purposes , because it combines various mathematical and physical principles. We developed a mathematical toy model for the time evolution of the buoyancy and the weight on such a tablet. According to this model we conclude at a mathematical formula which connects the lifting time of the tablet with the geometrical characteristics of it. We tested this formula experimentally and it seems to be confirmed pretty well, according to the experimental data.

**Introduction**
The effervescent tablets (vitamins, analgesic medicines, etc) are everywhere in our daily life.
Such solid tablet contains an active and others substances, which when it is dropped in a glass of water starts to react producing gas bubbles($CO_2$) and slowly breaks up disengaging the active substance in the water. The tablet, at first, goes at the bottom of the glass and gradually while its size shrinks, suddenly it goes up at a moment which we call lifting time.

The explanation of this effect from students(junior, high school, even from undergraduates) isn't always obvious.
The next hypothetical dialogue between a student and teacher summarize our observations about the main ideas students have for the effect.

*-Teacher; "Why does the tablet go at first at the bottom of the glass and after a while it starts to lift?"*

*-Student; "As the time pass and the tablet breaks up, its volume reduces so as a result it weighs less. At that moment, the buoyancy becomes bigger than the weight and it starts to lift."*

*-Teacher; "As the volume shrinks, the buoyancy becomes less so it's not obvious that it will be greater than the weight at some moment. What is more, if what you said was true, in case we cut a smaller piece from the tablet it should be floating immediately because of its smaller weight but this is not observable in practice."*

*-Student; "Maybe the bubbles that have been produced force it to lift, because they have smaller density from water. As the volume shrinks so does its weight and as a result at some moment the buoyancy will be bigger.*

*-Teacher: "While the volume and as a result the size of the tablet reduces, the number of bubbles and buoyancy does the same, so it's not obvious that the weight will be smaller than the buoyancy after a while."*
*-Student;…….*

We can observe that the students approaching against this effect, focuses on one parameter of the problem at a time, despite the fact that the explanation of this phenomenon requires a combination of physical and mathematical principles.

A qualitative explanation for this effect is that the weight is analog to the volume of the tablet, whereas the buoyancy depends mainly on the gas bubbles which adhere temporarily to the surface of the tablet.
Supposing that all the linear dimensions of the tablet are subjected to the same changes proportionally, the volume of the tablet reduces faster than the decrease of its surface, so at some point the buoyancy will be bigger than its weight. This is a widely known result of the geometry of solids that is used commonly.[1]

Consequently we will construct a mathematical toy model based on the main aforementioned principles.

**Mathematical toy model**
The assumptions we propound for this effect are:

a) All the linear dimensions of the tablet diminish proportionally the same.
b) The quantity of the bubbles created during the reaction with the water is analogous to the surface of the tablet.
c) Each bubble has the same volume.
d) The rate reaction of the mass of the tablet is analogous to the surface of the tablet.

Below we can see the symbols of the physical quantities we will use in our analysis.

$A$ ; The buoyancy on the tablet and the adhered bubbles on it.
$W$; The overall weight of the tablet and the bubbles
$V_{tb}$, $V_b$ ; The volume of the tablet and the mean volume of a bubble correspondingly
$V_{all,b}$ ; The overall volume of all the bubbles on the surface of the tablet.
$N_b$ ; The overall quantity of the adhered bubbles on the surface of the tablet
$S_{tb}$ ; The overall surface of the tablet
$m_{tb}$ ; The mass of the tablet
$d_{tb}$, $d_l$, $d_b$ ; The density of the tablet, water and the gas of the bubble correspondingly

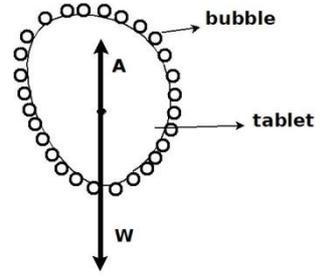

*figure 1. Buoyancy (A) due to the bubbles and weight( W) on an effervescent tablet.*

Due to assumption (b) we can set the surface density of the number of bubbles $sd_b = \frac{N_b}{S_{tb}}$ which is constant during the reaction and it only depends on the nature of the reactants.

The resultant force of weight and buoyancy(figure 1) which is exerted on the system of the tablet and the bubbles is;
$\Sigma F = A - W$

For A, W and $V_{all,b}$, applies that $A = (V_{all,b} + V_{tb}) \cdot d_l \cdot g$ , $W = V_{all,b} \cdot d_b \cdot g + V_{tb} \cdot d_{tb} \cdot g$
and $V_{all,b} = N_b \cdot V_b = sd_b \cdot S_{tb} \cdot V_b$

By replacing A and W with the resultant force we rich to the following result;
$\Sigma F = sd_b \cdot V_b \cdot (d_l - d_b) \cdot g \cdot S_{tb} - (d_{tb} - d_l) \cdot g \cdot V_{tb}$

The terms that multiply the surface $S_{tb}$ and the volume $V_{tb}$ are positive and constant during the reaction. If we set the constants $c_1 = sd_b \cdot V_b \cdot (d_l - d_b)$ and $c_2 = (d_{tb} - d_l) \cdot g$, we can rewrite the $\Sigma F$ as; $\Sigma F = c_1 \cdot S_{tb} - c_2 \cdot V_{tb}$ (1)

Based on assumption (a) and according to appendix 1 it must hold, for the surface and the volume of the tablet, that $S_{tb} = \left(\frac{V_{tb}}{C}\right)^{2/3}$ where $C = V_{tb0}/S_{tb0}^{3/2}$ and $V_{tb0}, S_{tb0}$ is the volume and the surface of the tablet at that moment when the reaction starts.

By replacing it to (1) we take; $\Sigma F = c_1 \cdot \left(\frac{V_{tb}}{C}\right)^{2/3} - c_2 \cdot V_{tb}$ => $\Sigma F = V_{tb}^{2/3} \cdot \left(\frac{c_1}{C^{2/3}} - c_2 \cdot V_{tb}^{1/3}\right)$ (2)

We can observe that as the volume reduces, the term inside of the parentheses is definitely becoming zero, so at that moment the tablet starts to lift.

Consequently, we will find how the volume changes as time passes. Based on assumption (d) we rich to the conclusion that $\frac{dm_{tb}}{dt} = -l \cdot S_{tb}$ (3), where l is a positive constant which depends on the nature of the reactants.

For the mass applies that $m_{tb} = d_{tb} \cdot V_{tb}$, so $\frac{d(d_{tb} \cdot V_{tb})}{dt} = -l \cdot S_{tb} \Rightarrow d_{tb} \cdot \frac{dV_{tb}}{dt} = \frac{-l}{C^{2/3}} \cdot V_{tb}^{2/3}$.

Assigning the constant $k = \frac{l}{d_{tb} \cdot C^{2/3}}$, we conclude that $\frac{dV_{tb}}{dt} = -k \cdot V_{tb}^{2/3}$ (4)

Considering that at $t_0 = 0$ the volume has value $V_{tb0}$, the solution of (4) (see appendix 3) is $V_{tb}(t) = \left(V_{tb0}^{1/3} - \frac{k}{3} \cdot t\right)^3$ (5)

Combining the equations (2) and (5) we can find the time progress of the $\Sigma F$.

A physical quantity we can measure easily is the lifting time $t_{lift}$ ; namely, the period of time at which the reaction starts, till the moment that the tablet starts to lift.

From equation (2) we observe that if at $t_0 = 0$ applies that $\frac{c_1}{C^{2/3}} < c_2 \cdot V_{tb0}^{1/3}$, the weight will come uppermost to the buoyancy. In this case, the tablet will go down.

The tablet will start the ascension at the moment $t_{lift}$ when the $\Sigma F$ becomes zero;
$\Sigma F(t_{lift}) = 0 \Rightarrow \frac{c_1}{C^{2/3}} - c_2 \cdot V_{tb}^{1/3} = 0 \Rightarrow V_{tb}^{\frac{1}{3}} = \frac{c_1}{c_2 \cdot C^{2/3}}$ (6)

Combining (5) and (6) we take ; $V_{tb0}^{\frac{1}{3}} - \frac{k}{3} \cdot t_{lift} = \frac{c_1}{c_2 \cdot C^{2/3}}$. From this equation and the definitions of the constants k, C

(having set the constants $A = \frac{3 \cdot d_{tb}}{l}$ and $B = \frac{3 \cdot d_{tb} \cdot c_1}{l \cdot c_2}$ ) we rich our final result for the lifting time $t_{lift} = A \cdot \frac{V_{tb0}}{S_{tb0}} - B$ **(7)**

At the moment when the reaction starts we observe that the lifting time depends linearly on the ratio of the volume to the surface of the tablet.
Consequently, we will test experimentally the validity of this conclusion.

**Experiment and Results**
In this experiment we examined the validity of the equation (7) , by dropping tablets of different sizes and ratio $V_{tb0}/S_{tb0}$ in a container with water and measuring the lifting time $t_{lift}$ for each tablet.

For our measurements we used a complete tablet ($\theta = 2\pi$) and tablets we cut ,with knife, into angles $\theta = \pi, \pi/2, \pi/4$ as it appears at figure 2.

For each different size, we measured up to five times the lifting time $t_{lift}$ and we calculated the mean value and the corresponding statistical error. The ratio $V_{tb0}/S_{tb0}$ (for its calculations see appendix 3) and the corresponding measurements are shown at the table above.

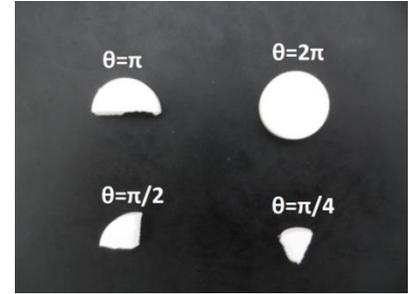

*figure 2. Tablets of different sizes that we used in the experiment.*

The radius (R) and the width (d) of the tablets we used are; $R = (11.56 \pm 0.01)mm$ and $d = (5,90 \pm 0.01)mm$

| θ(rad) | $V_{tb0}(mm^3)$ | $S_{tb0}(mm^2)$ | $\frac{V_{tb0}}{S_{tb0}}(mm)$ | $t_{lift1}(sec)$ | $t_{lift2}(sec)$ | $t_{lift3}(sec)$ | $t_{lift4}(sec)$ | $t_{lift5}(sec)$ | $\bar{t}_{lift}(sec)$ | $\delta t_{lift}(sec)$ |
|---|---|---|---|---|---|---|---|---|---|---|
| 2π | 2477.0 | 1268.2 | 1.95 | 52.0 | 50.4 | 65.4 | 69.1 | 66.3 | 60.6 | 3,9 |
| π | 1238.5 | 770.5 | 1.61 | 49.8 | 49.6 | 57.4 | 61.4 | 64.9 | 56.6 | 3,1 |
| π/2 | 619.2 | 453.5 | 1.37 | 30.3 | 47.2 | 47.9 | 48.4 | 56.0 | 46.0 | 4,2 |
| π/4 | 309.6 | 294.9 | 1.05 | 29.5 | 23.5 | 30.4 | - | - | 27.8 | 2,2 |

Table 1. Measurements of the lifting time relating with the ratio $V_{tb0}/S_{tb0}$. Due to difficulties cutting tablets that have angle θ=π/4 (they fall apart easily with the knife) we took measurements with three pieces only.

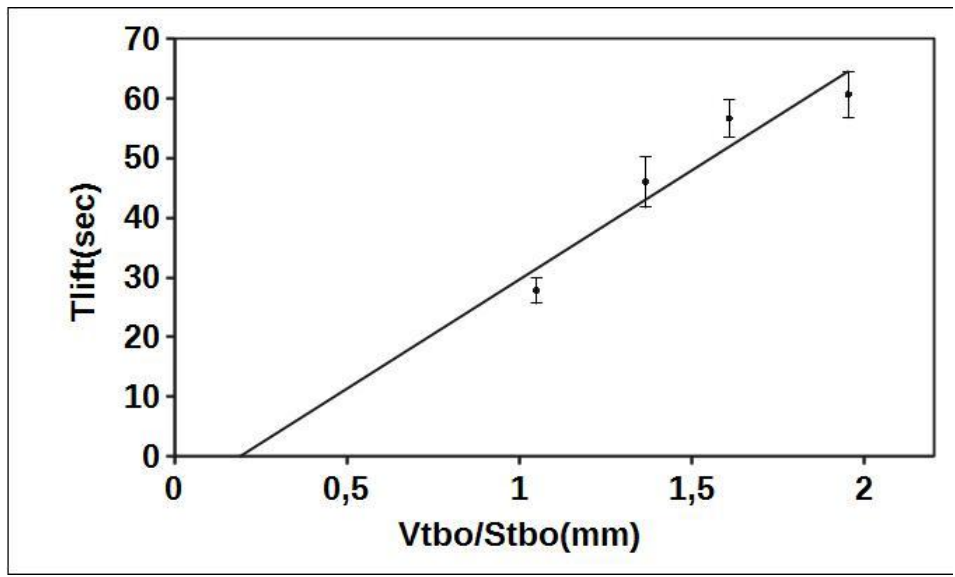

figure 3. Diagram showing the lifting time relating with the ratio Vtbo/Stbo. Also, it has been drawn the relative least square regression line.

From figure 3 we can see that the lifting time relating with the ratio $V_{tb0}/S_{tb0}$ in the range of the values that we examined can be fitted pretty well with a least square linear regression line. So, it seems that they have a linear dependence, as the mathematical model predicts.

The linear correlation coefficient has the value of R=0.95. The values of the coefficients A, B of the equation $t_{lift} = A \cdot \frac{V_{tb0}}{S_{tb0}} - B$ are; $A = 36.6 \frac{sec}{mm}$ and $B = 6.93 sec$

## Conclusions

The weight of an effervescent tablet depends on its volume and the buoyancy comes mainly from the overall buoyancy of its bubble that adhere on its surface during the reaction with the water. Considering that all the linear sizes of the tablet reduce analogously the same, we reach at the conclusion that, the buoyancy at some specific point becomes larger than the weight and as a result the tablet lifts.

Making use of a mathematical toy model we concluded that each lifting time depends linearly on the ratio of the volume to the surface that the tablet has at the point we drop it in the water.

A future experimental examination could be the time progress of the resultant force on the tablet and its comparison with the theoretical behavior which was expected.

The educational use of this phenomenon is that it contains;
i) the exploration of the appropriate natural principles and the construction of a theoretical model that interprets the effect.
ii) the experimental procedure for the investigation of a conclusion that emerge from the model.

We think that this activity could be integrated in a project for undergraduate or advanced high school students.

## Appendix
### 1) Dependence of volume - surface of an object when its linear dimensions changes analogously the same.

Let's assume that the volume and the surface of an object depend on the lengths $l_1, l_2, ....$, through homogeneous functions f, g of 3rd and 2nd degree respectively, namely, $V=f(l_1,l_2,..)$ and $S=g(l_1,l_2,..)$. Let's suppose that the lengths change the same with rate k, namely $l_i' = k \cdot l_i$

Then, for the new volume V' and the surface S' applies that;

$$V' = f(l_1', l_2', ...) = f(k \cdot l_1, k \cdot l_2, ..) = k^3 \cdot f(l_1, l_2, ..) = k^3 \cdot V \qquad (1)$$

$$S' = g(l_1', l_2', ...) = g(k \cdot l_1, k \cdot l_2, ..) = k^2 \cdot g(l_1, l_2, ..) = k^2 \cdot S \Rightarrow k = \left(\frac{S'}{S}\right)^{\frac{1}{2}} \qquad (2)$$

Combining equations (1), (2) we conclude that $V' = \left(\frac{S'}{S}\right)^{3/2} \cdot V \Rightarrow \frac{V'}{S'^{3/2}} = \frac{V}{S^{3/2}} = C = \text{constant}$

### 2) Volume and surface of the tablet
The shape of the tablets we used seems like the figure 4.

The volume and the surface in contrast with the shape are;

figure 4(a); $V = \pi R^2 d$, $S = 2\pi Rd + 2\pi R^2$

figure 4(b); $V = \frac{1}{2}\theta R^2 d$, $S = (2+\theta)Rd + \theta R^2$,

where the angle θ is measured in rad.

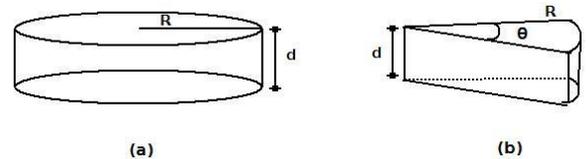

*figure 4. Geometry of the tablets we used in the experiment.*

### 3) Solution of the differential equation $\frac{dV_{tb}}{dt} = -k \cdot V_{tb}^{2/3}$

Considering that at $t_0 = 0$ the volume of the tablet has the value of $V_{tb0}$, namely $V_{tb}(0) = V_{tb0}$, then

$$\frac{dV_{tb}}{dt} = -k \cdot V_{tb}^{\frac{2}{3}} \Rightarrow dV_{tb} \cdot V_{tb}^{\frac{-2}{3}} = -k \cdot dt \Rightarrow \int_{V_{tb0}}^{V_{tb}} dx \cdot x^{\frac{-2}{3}} = -\int_0^t k\, dt \Rightarrow \left[3x^{\frac{1}{3}}\right]_{V_{tb0}}^{V_{tb}} = -k \cdot t \Rightarrow$$

$$\Rightarrow V_{tb}(t) = \left(V_{tb0}^{\frac{1}{3}} - \frac{k}{3} \cdot t\right)^3$$

## References
[1] John A. Adam, 2009, "A Mathematical Nature Walk", Princeton University Press, see for example, Q21 page 21